\documentstyle[preprint,tighten,prl,aps,floats]{revtex}

\begin{document}

\preprint{IFUP-TH 27/99}

\title{High-precision determination of the critical exponents for the
$\lambda$-transition of ${}^4$He by improved high-temperature
expansion}

\author{Massimo Campostrini$\,^1$, Andrea Pelissetto$\,^2$, 
        Paolo Rossi$\,^1$, Ettore Vicari$\,^1$ }
\address{$^1$ Dipartimento di Fisica dell'Universit\`a di Pisa 
and I.N.F.N., I-56126 Pisa, Italy}
\address{$^2$ Dipartimento di Fisica dell'Universit\`a di Roma I
and I.N.F.N., I-00185 Roma, Italy \\
{\bf e-mail: \rm
{\tt campo@mailbox.difi.unipi.it}, 
{\tt rossi@mailbox.difi.unipi.it},
{\tt Andrea.Pelissetto@roma1.infn.it},
{\tt vicari@mailbox.difi.unipi.it}
}}
\date{May 27, 1999}

\maketitle

\begin{abstract}
We determine the critical exponents for the XY universality class in
three dimensions, which is expected to describe the
$\lambda$-transition in ${}^4$He.  They are obtained from the analysis
of high-temperature series computed for a two-component
$\lambda\phi^4$ model. The parameter $\lambda$ is fixed such that the
leading corrections to scaling vanish. We obtain $\nu = 0.67166(55)$,
$\gamma = 1.3179(11)$, $\alpha=-0.0150(17)$. These estimates improve
previous theoretical determinations and agree with the more precise
experimental results for liquid Helium.
\end{abstract}

\vskip1.5pc
\bgroup\small
\leftskip=0.10753\textwidth \rightskip\leftskip
\par\egroup
\vskip1.5pc


The renormalization-group approach to cooperative transitions is one
of the most significant successes in theoretical physics.  It has
explained a wide range of phenomena in many different fields, ranging
from statistical physics to elementary particle physics. It should be
noted however that the quantitative experimental support for the
theory rests primarily on its predictions for critical phase
transitions.  To test the theory one needs the most accurate possible
values for universal quantities. Here the ${}^4$He phase transition is
of great utility because of the weakness of the singularity of the
compressibility of the fluid and of the availability of extremely pure
samples. Moreover the possibility of performing the experiments in
space, and therefore in a microgravity environment, reduces the
gravity-induced broadening of the transition.  Recently a Space
Shuttle experiment \cite{Lipa-etal_96} performed a very precise
measurement of the heat capacity of liquid helium to within 2 nK from
the $\lambda$-transition obtaining an extremely accurate estimate of
the exponent $\alpha$:
\begin{equation}
\alpha = - 0.01285\pm 0.00038.
\label{alpha-sperimentale}
\end{equation}
This estimate is extremely precise and represents a challenge 
for theorists, who, until now, have not been able to compute 
critical exponents at this level of accuracy. 

One of the oldest approaches to the study of critical phenomena is
based on high-temperature (HT) expansions. In this approach the main
hindrance to a precise determination of universal quantities is the
presence of confluent corrections with non-integer exponents. For
instance the specific heat is supposed to behave as
\begin{equation}
C_p = A t^{-\alpha}\left(1 + B t^\Delta + C t\ldots \right) + 
      D + E t + \ldots 
\label{Cp-espansione}
\end{equation}
for $t\equiv |T-T_c|/T_c\to 0$, with $\Delta\approx 0.5$.  The
presence of non-analytic terms introduces a large (and dangerously
undetectable) systematic error in the results of the HT series
analysis.  In order to obtain precise estimates of the critical
parameters, the approximants of the HT series should properly allow
for the confluent non-analytic corrections.  Integral (also called
differential) approximants \cite{int-appr-ref} are, in principle, able
to describe a behaviour of the type (\ref{Cp-espansione}) (see e.g.\ 
Ref.\ \cite{Guttrev} for a review).  However, the extensive numerical
work that has been done for the Ising model shows that in practice,
with the series of moderate length that are available today, no
analysis is able to predict and take into account non-analytic
correction-to-scaling terms
\cite{ZJ-79,Nickel-82,Nickel-Dixon_81,C-F-N-82,Adler-83,N-R-90}.  In
order to effectively keep into account these confluent corrections,
one should use biased approximants, fixing the value of $\beta_c$ and
of the first non-analytic exponent $\Delta$ (see e.g.\ Refs.\ 
\cite{Roskies-81,Nickel-Dixon_81,A-M-P-82,B-C-97,P-V-gr-98,B-C-g-98}).

To overcome these difficulties, in the early 80s, Chen, Fisher, and
Nickel \cite{C-F-N-82} realized the importance of studying families of
models (specified by some auxiliary parameter) which are candidates
for belonging to the same universality class. The hope was the
possibility of locating a parameter value at which the leading
non-analytic corrections vanish.  If the leading non-analytic terms
are no longer present, one expects a faster convergence, and therefore
more precise and reliable estimates of the critical quantities.  The
method was applied to the double-Gaussian and to the Klauder models,
both belonging to the Ising universality class and depending
continuously on a real parameter; it was shown that a Hamiltonian for
which the leading corrections are suppressed --- we will name it
``improved" Hamiltonian --- could indeed be found
\cite{C-F-N-82,G-R-84,F-C-85}. The crux of the method is the precise
determination of the optimal value of the parameter appearing in the
Hamiltonian.  In Refs.\ \cite{C-F-N-82,G-R-84} the partial
differential approximant technique was used; however the errors on the
improved Hamiltonian were relatively large and the final results
represented only a modest improvement with respect to standard (and
much simpler) analyses using biased approximants.

In the last few years it has been understood that improved
Hamiltonians can be determined with high accuracy by means of Monte
Carlo simulations
\cite{H-P-V-98,B-F-M-M-98,B-F-M-M-P-R-99,Hasenbusch-99}.  Using
finite-size scaling methods that are very sensitive to confluent
corrections, it has been possible to determine various improved
Hamiltonians that belong to the Ising universality class, and
correspondingly precise estimates of critical quantities have been
obtained.  It is important to notice that the estimates obtained using
different improved Hamiltonians agree within the quoted error bars,
confirming the correctness of the error estimates.  Similar methods
have been applied to the determination of critical exponents for
dilute polymers in good solvents \cite{B-N-97}.

The possibility of determining precisely the ``improved" value of the
parameter for a family of Hamiltonians has recently revived the
program of Ref.\ \cite{C-F-N-82}. In Ref.\ \cite{Campostrini-etal_99}
we performed an extensive analysis for the Ising universality class.
Critical exponents and many other universal quantities were determined
by analyzing HT series for three different improved Hamiltonians. For
each improved model, we obtained very accurate estimates
\cite{esponenti-Ising-3d} that were in good agreement among each
other, confirming the correctness of the quoted errors. This analysis
showed that, once a precise estimate of the improved parameter is
available, the HT series analysis gives results of a quality
comparable with or better than the best Monte Carlo simulations.

As is well known, see e.g. Ref.\ \cite{Buckingham}, 
the $\lambda$-transition of liquid Helium 
is expected to be in the XY universality class.
In order to check if this is really the case, 
it is important to have precise theoretical estimates that 
can be compared with the experimental results. 
Recently, Hasenbusch and T\"or\"ok \cite{Hasenbusch-Toeroek_99} 
performed a high-precision simulation of the $O(2)$
$\phi^4$ model, obtaining an accurate 
estimate of the exponent $\nu$: $\nu = 0.6723(3)(8)$. 
Using the hyperscaling relation $\alpha = 2 - 3 \nu$, 
they obtained 
\begin{equation}
\alpha = -0.0169\pm 0.0033.
\label{alpha-Hasenbusch-Toeroek}
\end{equation}
This estimate is larger than the experimental result
(\ref{alpha-sperimentale}), although the difference is barely bigger
than the error.  It is therefore important to further improve the
accuracy of the theoretical estimates in order to understand whether
this small discrepancy is significant.  For our purposes Ref.\ 
\cite{Hasenbusch-Toeroek_99} is particularly relevant since it
provides a quite precise determination of an improved Hamiltonian
belonging to the XY universality class.  In this paper we study this
improved model and obtain very precise estimates of the critical
exponents from the HT series analysis.  They considerably improve
previous HT estimates, showing the effectiveness of the approach.  In
particular we obtain for $\alpha$
\begin{equation}
\alpha = - 0.0150 \pm 0.0017, 
\label{alpha-nostro}
\end{equation}
which is in better agreement with the experimental result, although
still slightly larger. The small discrepancy between the Monte Carlo
and the experimental result is significantly reduced, providing
support to the fact that the $\lambda$-transition belongs to the same
universality class of the XY model. The full set of estimates together
with other recent results is reported in Table \ref{table_exponents}.

\begin{table}
\begin{center}
\begin{tabular}{lllll}
\hline
 & $\gamma$ & $\nu$ & $\eta$ & $\alpha$  \\
\hline \hline
IHT[this work] & 1.3179(7+4) & 0.67166(33+22)& 0.0381(2+1) & 
                $-$0.0150(10+7)$^*$\\
\hline
HT, sc \protect\cite{B-C-97}  & 1.325(3) & 0.675(2) & 0.037(7)$^*$ & 
                  $-$0.025(6)$^*$ \\
HT, bcc \protect\cite{B-C-97} & 1.322(3) & 0.674(2) & 0.039(7)$^*$ & 
                  $-$0.022(6)$^*$ \\
HT \protect\cite{B-C-99} & & & & $-$0.014(9), $-$0.022(6) \\
\hline
MC \protect\cite{Hasenbusch-Toeroek_99} & 1.319(2)$^*$ 
              & 0.6723(3)(8) & 0.0381(2)(2) & $-$0.0169(33)$^*$ \\
MC \protect\cite{Ballesteros-etal_96}   & 1.316(3)$^*$
              & 0.6721(13)  & 0.0424(25) &  $-$0.0163(39)$^*$ \\
MC \protect\cite{K-L-99}   & 1.315(12)
              & 0.6693(58)  & 0.035(5) &  $-$0.008(17)$^*$ \\
MC \protect\cite{Hasenbusch-Gottlob_93} & 1.307(14)$^*$
              & 0.662(7)    & 0.026(6) &  $-$0.014(21)$^*$ \\
MC \protect\cite{Janke_90}              & 1.323(2) 
              & 0.670(2)    & 0.025(7)$^*$ & $-$0.010(6)$^*$ \\
\hline
$d=3$ $g$-exp \protect\cite{Guida-ZinnJustin_98} 
                 & 1.3169(20) & 0.6703(15) & 0.0354(25) & $-$0.011(4)$^*$ \\
$\epsilon$-exp, free \protect\cite{Guida-ZinnJustin_98}
                 & 1.3110(70) & 0.6680(35) & 0.0380(50) & $-$0.004(11)$^*$ \\
$\epsilon$-exp, bc \protect\cite{Guida-ZinnJustin_98}
                 & 1.317      & 0.671      & 0.0370     & $-$0.013$^*$    \\
\hline
${}^4$He \cite{Lipa-etal_96}    & & 0.67095(13)$^*$ & & $-$0.01285(38) \\
${}^4$He \cite{Goldner-etal_93} & & 0.6705(6)     & &   $-$0.0115(18)$^*$ \\
${}^4$He \cite{Swanson-etal_92} & & 0.6708(4)     & &   $-$0.0124(12)$^*$ \\
\hline
\end{tabular}
\end{center}
\caption{Estimates of the critical exponents.
See text for the explanation of the symbols in the first column.
We indicate with an asterisk (${}^*$) the estimates that have 
obtained using the hyperscaling relation $2 - \alpha - 3 \nu = 0$
or the scaling relation $\gamma - (2 - \eta)\nu = 0$.  bcc and sc 
refer to the body-centered cubic and to the simple cubic lattices
respectively.
 }
\label{table_exponents}
\end{table}

We consider a simple cubic lattice and the Hamiltonian
\begin{equation}
{\cal H}=\, - \beta\sum_{<x,y>} {\vec\phi}_x\cdot{\vec\phi}_y +\, 
         \sum_x \left[ {\vec\phi}_x^2 + \lambda ({\vec\phi}_x^2 - 1)^2\right],
\label{Hamiltonian}
\end{equation}
where $<x,y>$ labels a lattice link, and ${\vec\phi}_x$ is a real
two-component vector defined on lattice sites.  Hasenbusch and
T\"or\"ok have performed an extensive simulation of this model and
have obtained an estimate of the value of $\lambda$, $\lambda^*$, at
which the leading corrections vanish.  They quote:
\begin{equation}
\lambda^* =\, 2.10 \pm 0.01 \pm 0.05,
\label{stimalambdastar}
\end{equation}
where the two errors are respectively the statistical and the 
systematic one.  We have considered the susceptibility $\chi$ and the 
second-moment correlation length $\xi$ defined by
\begin{eqnarray}
\chi &=& \sum_x \langle {\vec\phi}_0\cdot{\vec\phi}_x\rangle, \\
\xi  &=& {1\over 6\chi} \sum_x |x|^2 
         \langle {\vec\phi}_0\cdot{\vec\phi}_x\rangle.
\end{eqnarray}
Using the linked-cluster expansion technique, we have generated the HT
expansion of these two quantities to 20th order for arbitrary values
of $\lambda$. From the analysis of $\chi$ and $\xi$ we can obtain
directly estimates of $\gamma$ and $\nu$. In the analysis we used
inhomogeneous integral approximants (IA).  Second- and third-order
IA's turned out to be the most stable, especially when we biased the
approximants \cite{F-C-85} requiring the presence of two symmetric
singularities at $\beta=\, \pm\beta_c$.  This is a natural
requirement, since it can be proved that, on bipartite lattices,
$\beta=-\beta_c$ is also a singular point associated to the
antiferromagnetic critical behavior \cite{Fisher-62}.  The results we
quote were obtained using this type of approximants.

In Table \ref{table_exponents} we give our estimates of the exponents
$\gamma$ and $\nu$ that have been obtained analyzing the HT series for
the improved Hamiltonian (\ref{Hamiltonian}) with $\lambda=\lambda^*$.
We quote two errors: the first one is related to the spread of the
approximants, while the second one gives the variation of the exponent
when $\lambda$ varies between 2.04 and 2.16, cf.\ Eq.\ 
(\ref{stimalambdastar}).  It should be noted that the first error is
somewhat larger than the second one, and therefore it is important to
further extend the HT series. However, since the second error is far
from negligible, a substantial reduction of the uncertainty also
requires a more accurate determination of $\lambda^*$.  Note that this
might also reduce the first error, since it would enable us to work
closer to the exactly improved Hamiltonian. We do not present details
on the generation and analysis of the HT series and we refer the
reader to App.\ A of Ref.\ \cite{Campostrini-etal_99}.

From the estimate of $\nu$ we can obtain the exponent $\alpha$
assuming the validity of the hyperscaling relation
\begin{equation}
\alpha = 2 - 3 \nu.
\label{hyper}
\end{equation}
In principle it should be possible to determine $\alpha$ directly from
the specific heat, or from the singularity of the susceptibility at
the antiferromagnetic point \cite{Fisher-62}. We tried this second
method obtaining only a rough estimate: $\alpha=-0.02(2)$.

From the estimates of $\gamma$ and $\nu$
it is possible to obtain the exponent $\eta$, using 
\begin{equation}
\gamma = \nu(2 - \eta).
\label{gammasunu}
\end{equation}
However, it is not clear how to set the error bar on the result.  One
can use the independent-error formula taking into account the error on
$\gamma$ and $\nu$, but this may be an overestimate since $\gamma$ and
$\nu$ are correlated. To obtain an estimate of $\eta$ with a smaller,
yet reliable, error bar, we used the so-called critical-point
renormalization method (see e.g. \cite{int-appr-ref} and references
therein).  The value that is quoted in Table \ref{table_exponents} has
been obtained with this method.  It is compatible with the estimate
obtained using the scaling relation (\ref{gammasunu}), but it has a
smaller error bar.

In Table \ref{table_exponents} we report a summary of the most precise
estimates that have been obtained in the last years.  When only $\nu$
or $\alpha$ was reported, we used the relation (\ref{hyper}) to obtain
the missing exponent.  Analogously if only $\eta$ or $\gamma$ was
quoted, the second exponent was obtained using the scaling relation
(\ref{gammasunu}); in this case the uncertainty was obtained using the
independent-error formula. The results we quote have been obtained
from the analysis of the HT series of the XY model (HT), by Monte
Carlo simulations (MC) or by field-theory methods.  The field-theory
results have been derived by resumming the perturbative expansion in
fixed dimension $d=3$ ($g$-expansion), or the the expansion in
$\epsilon=4-d$.  For the $\epsilon$-expansion we quote two numbers,
corresponding to an unconstrained analysis (``free"), and to a
constrained analysis (``bc") in which the two-dimensional values of
the exponents are taken into account.  Our final results for $\gamma$
and $\nu$ improve the existing theoretical estimates. They are
somewhat lower than previous HT results but they are in full agreement
with the field-theory and the most recent MC estimates, as well as
with the experimental results for the $\lambda$-transition, showing
the expected universality.


\begin{thebibliography}{99}
\bibitem{esponenti-Ising-3d}
For the Ising universality class, we obtained:
$\gamma = 1.2373(4)$, $\nu=0.63015(25)$, and $\eta = 0.0364(4)$.
\end{thebibliography}
\end{document}